\documentclass[11pt,a4paper]{article}

\usepackage{amsmath,amsthm,amssymb}
\usepackage{graphics,graphicx}
\usepackage{epsfig}
\usepackage{multicol}
\usepackage{color}
\makeatletter
\@addtoreset{equation}{section}
\makeatother

\DeclareMathAlphabet{\mathpzc}{OT1}{pzc}{m}{it}
\begin{document}

%\begin{flushright}
%{RESCEU-??/12}
%\end{flushright}

\begin{center}
\LARGE{\bf Lumpy cosmic strings}
\end{center}

\begin{center}
\large{\bf Matthew J. Lake} ${}^{a,b}$\footnote{matthewj@nu.ac.th}\large{\bf and Tiberiu Harko} ${}^{c}$\footnote{t.harko@ucl.ac.uk}
\end{center}
\begin{center}
\emph{ ${}^a$ The Institute for Fundamental Study, ``The Tah Poe Academia Institute", \\
Naresuan University, Phitsanulok 65000, Thailand and \\}
\emph{ ${}^b$ Thailand Center of Excellence in Physics, Ministry of Education, Bangkok 10400, Thailand}
\emph{ ${}^c$ Department of Mathematics, University College London, \\
Gower Street, London WC1E 6BT, United Kingdom \\}
\vspace{0.1cm}
\end{center}

%%%%%%%%%%%%%%%%%%%%%%%%%%%%%%%%%%%%%%%%%%%%%%%%%%%%%%%%%%%%%%%%
%%%%%%%%%%%%%%%%%%%%%%%%%%%%%%%%%%%%%%%%%%%%%%%%%%%%%%%%%%%%%%%%``

\begin{abstract}
We outline a model of abelian-Higgs strings with variable scalar and vector core radii. In general, the functions determining the time and position-dependent core widths may be expressed as arbitrary left or right movers, of which the usual constant values are a particular solution. In this case the string may carry momentum, even if the embedding of its central axis remains fixed, and the resulting objects resemble ``necklaces". Some possible astrophysical applications of lumpy strings, including as potential engines for anomalous gamma ray bursts, are also discussed.
\end{abstract}

%%%%%%%%%%%%%%%%%%%%%%%%%%%%%%%%%%%%%%%%%%%%%%%%%%%%%%%%%%%%%%%%%``
%%%%%%%%%%%%%%%%%%%%%%%%%%%%%%%%%%%%%%%%%%%%%%%%%%%%%%%%%%%%%%%%%
%
\section{Introduction} \label{Sect.I}
In 1973, Nielsen and Olesen's seminal paper, demonstrating the existence of a topologically stabilized vortex solution in the abelian-Higgs model \cite{NiOl:73}, marked the beginning of the theory of cosmic strings as topological defects, formed during symmetry breaking phase transitions in the early universe \cite{ViSh:00,HiKi:99}. Since then, an enormous number of string-type solutions have been found in many different field-theoretic models, including electroweak strings in the Weinberg-Salam model \cite{Vachaspati:1992uz,James:1992zp,James:1992wb}, and strings in GUT scale theories \cite{Kakushadze:1997mc}. Phenomenologically, superconducting strings are also of great interest in cosmology and astrophysics \cite{W,Nielsen:1987fy}.\\
\indent
Interest in cosmic strings was also renewed after the discovery that $F$-strings (fundamental strings governed by the Nambu-Goto action) and $D$-strings (one-dimensional $D$-branes), could be formed in large numbers at the end of inflation, in string theory inspired models of the early universe \cite{Polchinski:2004ia,Jones:2003da}. Though not field-theoretic, such strings are nonetheless ``cosmic" in origin and provide an alternative paradigm in which to study the cosmological implications of the existence of string-like objects. \\
\indent
With the exception of ``necklaces" \cite{Berezinsky:1997kd}-\cite{Thesis} which, in many models, are a form of hybrid defect; that is, a series of monopole-like ``beads" connected by strings, rather than strings in their own right, virtually all these models have one thing in common. Whatever the underlying nature of the string, it is, essentially, a ``one-dimensional" object. Though, strictly, this is only true for cosmic superstrings, at the classical level, while topological defect strings necessarily occupy a three-dimensional volume, in the latter the string radius is assumed to be constant and much smaller than the string length, so that it may be neglected on cosmological scales. Even in the case of necklaces, the monopoles do not typically occupy large volumes, or have widths much larger than the string itself.  \\
\indent
An alternative model was proposed in \cite{Lake:2010wt}, in which a defect string with varying local radii was found to exist. By modifying the standard abelian-Higgs model to include spatially-dependent couplings for the scalar and vector fields, static, non-cylindrically symmetric solutions of the resulting field equations were obtained. Though phenomenologically interesting, in the absence of any conclusive proof for the spatial (or temporal) variation of fundamental constants \cite{Uzan:2002vq,Flambaum:2007my,Flambaum:2008ps}, such solutions must be considered as toy models. \\
\indent
In this paper, we propose a new solution of the standard abelian-Higgs field equations, which represents a generalization of the Nielsen-Olesen string. For a string whose central axis lies along the coordinate $z$-axis, this solution is characterized by the addition of two arbitrary functions of $z \pm t$, which represent arbitrary left or right movers, and which determine the effective local radii of the scalar and vector cores. The generalized solution thus represents a ``lumpy" cosmic string, and it is found that the lumps propagate at the speed of light. Phenomenologically, the solution combines several features of existing models in an interesting way. Specifically, it is non-cylindrically symmetric, like the static solution presented in \cite{Lake:2010wt}, but possesses momentum, as well as mass-energy, which may be interpreted as a kind of neutral current, as in superconducting strings. In addition, for lumps that are highly localized along the string length, it may resemble a necklace, with highly energetic traveling beads. \\
\indent
Importantly, however, unlike necklace models, there is no theoretical maximum value for the radius of the lumps, at least not in the classical theory considered here. In principle, quantum effects may limit its value to the same order of magnitude as the standard string width, though further detailed analysis is required to establish whether this indeed the case. Therefore, in order for lumpy strings with conserved momentum to form, all that is required is for local variations in the vortex radii, at the epoch of string formation, to be correlated, rather than random, at least over some characteristic length scale. Again, only a detailed and, possibly, quantum mechanical analysis of the field evolution during the symmetry breaking phase transition can establish whether this is possible on scales that will effect the macroscopic string dynamics. However, we assume, for now, that such correlations can exist, and determine the general features of the resulting lumpy string solution. \\
\indent
The structure of this paper is then as follows. In Secs. \ref{Sect.IIA} and \ref{Sect.IIB}, we review the abelian-Higgs model and the Nielsen-Olesen solution, respectively, while in Sec. \ref{Sect.IIIA} we present the lumpy string ansatz and demonstrate that the field equations may be solved consistently. In Sec. \ref{Sect.IIIB}, we determine the lumpy string energy and momentum, as well as the integrated values of the local pressures and shears. This allows us to estimate the energy and momentum emission as the string relaxes to the (lower energy) Nielsen-Olesen configuration, in Sec. \ref{Sect.IVA}, and some possible astrophysical implications of this are considered in Sec. \ref{Sect.IVB}. In particular, we consider the possibilities that energy emission from lumpy strings may provide an explanation for anomalous gamma ray bursts \cite{Kul}, or contribute to the observed positron excess coming from the galactic centre \cite{Fer} Finally, a brief discussion and summary of our results is given in Sec. \ref{Sect.V}.
%
%%%%%%%%%%%%%%%%%%%%%%%%%%%%%%%%%%%%%%%%%%%%%%%%%%%%%%%%%%%%%%%%%``
\section{Background: The abelian-Higgs model and the Nielsen-Olesen string} \label{Sect.II}
%
%%%%%%%%%%%%%%%%%%%%%%%%%%%%%%%%%%%%%%%%%%%%%%%%%%%%%%%%%%%%%%%%%``
\subsection{Recap of the abelian-Higgs model} \label{Sect.IIA}
In natural units, and using the metric convention $(+---)$, the abelian-Higgs action is
\begin{eqnarray} \label{AH_Act}
S = \int d^4x \sqrt{-g} \left\{D_{\mu}\phi \overline{D}^{\mu}\overline{\phi} - \frac{1}{4}F_{\mu\nu}F^{\mu\nu} - V(|\phi|)\right\},
\end{eqnarray}
where $\mu, \nu \in \left\{0,1,2,3\right\}$ and $V(|\phi|)$ is the potential term, which is determined by the symmetry breaking energy scale, $\eta$, and the scalar coupling, $\lambda$:
\begin{eqnarray} \label{AH_Pot}
V(|\phi|) = \frac{\lambda}{4}(|\phi|^2-\eta^2)^2.
\end{eqnarray}
Following the conventions of Vilenkin and Shellard \cite{ViSh:00}, the gauge covariant derivative $D_{\mu}$ and the electromagnetic field tensor $F_{\mu\nu}$, are defined according to
\begin{eqnarray} \label{Cov_Deriv}
D_{\mu} = \partial_{\mu} - ieA_{\mu}, \ \ \ F_{\mu\nu} = \partial_{\mu}A_{\nu} - \partial_{\nu}A_{\mu},
\end{eqnarray}
where $e$ is the vector coupling, and the covariant equations of motion (EOM) are
\begin{eqnarray} \label{Cov_EOMs}
\frac{1}{\sqrt{-g}}D_{\mu}\left(\sqrt{-g}D^{\mu}\phi\right) + \lambda\phi\left(|\phi|^2-\eta^2\right) = 0,
\end{eqnarray}
\begin{eqnarray} \label{CovEOMv}
\frac{1}{\sqrt{-g}}\partial_{\nu}\left(\sqrt{-g}F^{\mu\nu}\right) + ie\left(\overline{\phi}D^{\mu}\phi-\phi\overline{D}^{\mu}\overline{\phi}\right) = 0.
\end{eqnarray}
The action, Eq. (\ref{AH_Act}), is invariant under local $U(1)$ transformations of the form
\begin{eqnarray} \label{U(1)_invar}
\phi \rightarrow \phi' = \phi e^{i\Lambda(x)},Ê\ \ \ A_{\mu} \rightarrow A_{\mu}' = A_{\mu} + \frac{1}{e}\partial_{\mu}\Lambda(x),
\end{eqnarray}
where $\Lambda(x)$ is any single-valued real function, giving rise to the conserved current
\begin{eqnarray} \label{Cons_current}
j^{\mu} =  -ie\left(\overline{\phi}D^{\mu}\phi-\phi\overline{D}^{\mu}\overline{\phi}\right).
\end{eqnarray}
The conservation of $j^{\mu}$ is expressed directly via the vector EOM, Eq. (\ref{CovEOMv}), and the corresponding conserved charge is
\begin{eqnarray} \label{Q}
Q = \int j^{0} \sqrt{-g} d^3x.
\end{eqnarray}
The energy-momentum tensor is defined implicitly by varying the action with respect to the metric,
\begin{eqnarray} \label{T_mu_nu*}
\delta S = \int T^{\mu\nu} \delta g_{\mu\nu} \sqrt{-g}d^4x,
\end{eqnarray}
so that 
\begin{eqnarray} \label{T_mu_nu}
T^{\mu\nu} = \frac{-2}{\sqrt{-g}}\frac{\partial (\mathcal{L}\sqrt{-g})}{\partial g_{\mu\nu}} = D^{\mu}\phi\overline{D}^{\nu}\overline{\phi} + \overline{D}^{\mu}\overline{\phi}D^{\nu}\phi - F^{\mu}{}_{\alpha}F^{\nu\alpha} - g^{\mu\nu}\mathcal{L},
\end{eqnarray}
where  $\mathcal{L}$ is the Lagrangian density; that is, the sum of terms inside the curly brackets in Eq. (\ref{AH_Act}).
%
%%%%%%%%%%%%%%%%%%%%%%%%%%%%%%%%%%%%%%%%%%%%%%%%%%%%%%%%%%%%%%%%%``
\subsection{Recap of the Nielsen-Olesen solution} \label{Sect.IIB}
In cylindrical polar coordinates $(t,r,\theta,z)$ and assuming a Minkowski background so that,
\begin{eqnarray} \label{Metric1}
ds^2 = \eta_{\mu\nu}dx^{\mu}dx^{\nu} = c^2dt^2 - dr^2 - r^2d\theta^2 - dz^2,
\end{eqnarray}
the ansatz for the Nielsen-Olesen string is \cite{NiOl:73}
\begin{eqnarray} \label{NO_ansatz}
\phi(r,\theta) = \eta f(r) e^{in\theta}, \ \ \ A_{\theta} = \frac{n}{e}\alpha_{\theta}(r), \ \ \ A_r=A_z=A_0 = 0,
\end{eqnarray}
where $f(r)$ and $\alpha_{\theta}(r)$ are dimensionless functions which obey the boundary conditions
\begin{equation} \label{NO_f_bc}
f(r) = \left \lbrace
\begin{array}{rl}
0,& \ r=0 \\
1,& \ r \rightarrow \infty,
\end{array}\right.
\end{equation}
\begin{equation} \label{NO_a_{theta}_bc}
\alpha_{\theta}(r) = \left \lbrace
\begin{array}{rl}
0,& \ r=0 \\
1,& \ r \rightarrow \infty.
\end{array}\right.
\end{equation}
These give rise to the specific form of the EOM \cite{ViSh:00}
\begin{eqnarray} \label{NO_specific_scalar_EOM}
\frac{d^2 f}{d R_{s|n|}^2} + \frac{1}{R_{s|n|}} \frac{d f}{d r} - \frac{n^2 f}{R_{s|n|}^2}(1-\alpha_{\theta})^2 - \frac{1}{2}\left(\frac{r_{s|n|}}{r_s}\right)^2f(f^2-1) = 0,
\end{eqnarray}
\begin{eqnarray} \label{NO_specific_vector_EOM_theta}
\frac{d^2 \alpha_{\theta}}{d R_{v|n|}^2} - \frac{1}{R_{v|n|}} \frac{\partial \alpha_{\theta}}{\partial R_{v|n|}} + 2\left(\frac{r_{v|n|}}{r_v}\right)^2f^2(1-\alpha_{\theta}) = 0,
\end{eqnarray}
where we have defined
\begin{eqnarray} \label{R}
R_{i|n|} = \frac{r}{r_{i|n|}}, \ i \in \left\{s,v\right\},
\end{eqnarray}
and where $r_{s|n|}$ and $r_{v|n|}$ denote the values of the scalar and vector core radii for an $|n|$-vortex string, respectively.
We also use the simplified notation, $r_s$ and $r_v$, to refer to the scalar and vector core radii of $|n|=1$ strings and define the parameters
\begin{eqnarray}
\beta_{|n|} = \frac{r_{v|n|}^2}{r_{s|n|}^2}, \ \ \ \beta = \frac{r_{v}^2}{r_{s}^2},
\end{eqnarray}
for later convenience. Eqs. (\ref{NO_specific_scalar_EOM})-(\ref{NO_specific_vector_EOM_theta}) are solved, to leading order in the uncoupled regime, by \cite{HiKi:99,Thesis}
\begin{equation} \label{NO_f}
f(r) \approx \left \lbrace
\begin{array}{rl}
(r/r_{s|n|})^{|n|},& \ r \lesssim r_{s|n|} \\
1,& \ r \gtrsim r_{s|n|},
\end{array}\right.
\end{equation}
\begin{equation} \label{NO_alpha}
\alpha_{\theta}(r) \approx \left \lbrace
\begin{array}{rl}
(r/r_{v|n|})^{2},& \ r \lesssim r_{v|n|} \\
1,& \ r \gtrsim r_{v|n|}.
\end{array}\right.
\end{equation}
Strictly speaking, the power law solutions for $f(r)$ and $\alpha_{\theta}(r)$ given above are only valid in the ranges $r \ll r_{i|n|}$, $i \in \left\{s,v\right\}$, while the asymptotic forms hold only for $r \gg r_{i|n|}$.
However, by assuming that each holds, approximately, up to the limiting value $r = r_{i|n|}$, and ignoring the discontinuity of the derivatives, we can easily obtain order of magnitude estimates for the physical parameters of the string without the need for detailed numerical calculations.\\
\indent
The topological winding number, $n \in \mathbb{Z}$, is obtained as
\begin{eqnarray} \label{core_radii}
n = \frac{1}{2\pi}\int_{0}^{2\pi} \frac{\partial \vartheta}{\partial\theta}d\theta,
\end{eqnarray}
where $\vartheta$ represents the phase of the scalar field $\phi$. In general, the core radii are expected to be of the order of the Compton wavelengths of the associated bosons, $m_s^{-1}$ and $m_v^{-1}$, but may contain some $|n|$-dependence. This dependence is expected to take a simple form \cite{Thesis,Bo:76A,Bo:76B} so that here we assume
\begin{eqnarray} \label{core_radii}
r_{s|n|} &\approx& |n|^{\xi} m_s^{-1} = |n|^{\xi} (\sqrt{\lambda}\eta)^{-1},
\nonumber\\
r_{v|n|} &\approx& |n|^{\epsilon} m_v^{-1} = |n|^{\epsilon} (e \eta)^{-1},
\end{eqnarray}
where $\xi \geq 0$, $\epsilon \geq 0$ are constants.\\ 
\indent
Plugging the ansatz Eq. (\ref{NO_ansatz}), together with the approximate solutions in Eqs. (\ref{NO_f})-(\ref{NO_alpha}), into Eq. (\ref{T_mu_nu}), the only nonzero component of the energy-momentum tensor is $T^{00}$, so that the only constant of motion is the Hamiltonian, $E_{|n|}$, which depends on the absolute value of the topological winding number.
The (constant) mass-energy per unit length of the string, $\mu_{|n|} = \int T^{00}r dr d\theta$, is simply the total energy divided by the total length and, for $r_{s|n|}$ and $r_{v|n|}$ given by Eq. (\ref{core_radii}), it is straight forward to show that, assuming $r_v \geq r_s$ ($\beta \geq 1$), which corresponds to a Type II superconducting regime \cite{ViSh:00},
%\footnote{Strictly speaking, for core radii given by Eq. (\ref{core_radii}), we must assume $r_{v|n|} \geq r_{s|n|}$ ($\beta_{|n|} \geq 1$).}
\begin{eqnarray} \label{NO_mu2}
\mu_{|n|} \approx 2\pi \eta^2 \left[|n| + |n|^{2-2\epsilon} + |n|^{2\xi} + |n|^2\ln\left(\sqrt{\beta_{|n|}}\right)\right].
\end{eqnarray}
%at critical coupling ($e =\sqrt{\lambda}$, $\beta=1$);
%\begin{eqnarray} \label{NO_mu2}
%\mu_{|n|} \approx 2\pi \eta^2 (|n|^{2\sigma} + |n|^{2\eta}).
%\end{eqnarray}
However, in pioneering work \cite{Bo:76A,Bo:76B}, Bogomol'nyi showed that, for a Nielsen-Olesen string at critical coupling ($e = \sqrt{\lambda}$, $\beta = 1$),  $\mu_{|n|} $ must satisfy the condition
\begin{eqnarray} \label{NO_mu2}
\mu_{|n|} \geq 2\pi \eta^2 |n|,
\end{eqnarray}
otherwise known as the Bogomol'nyi-Prasad-Sommerfield, or BPS bound \cite{PrSo:75}. Since stability implies saturation of the bound \cite{ViSh:00} (see also \cite{deVe:76} for further work on the stability of abelian-Higgs strings), we require
\begin{eqnarray} \label{ParamBounds1}
0 \leq \xi \leq 1/2, \ \ \  \epsilon \geq 1/2,
\end{eqnarray}
for $|n| \sim \mathcal{O}(1)$. Alternatively, we may set
\begin{eqnarray} \label{ParamBounds2}
\xi = \epsilon = 1/2,
\end{eqnarray}
for arbitrary $|n|$. It is common in the literature to take $\sigma=0$ and $\epsilon = 1/2$, following Bogomol'nyi's original assumptions \cite{Bo:76A,Bo:76B}, which satisfy the conditions in Eq. (\ref{ParamBounds1}), but in the present work we adopt the latter conditions, given in Eq.  (\ref{ParamBounds2}), to allow for large $|n|$. For any applicable choice of parameters, the order of magnitude estimate for the mass-energy per unit length of a Nielsen-Olesen string with winding number $n$, at critical coupling, is
\begin{eqnarray} \label{NO_mu3}
\mu_{|n|} \approx 2\pi \eta^2 |n|.
\end{eqnarray}
Though it is not physically necessary for the components of the total tension vector within the string $\mathcal{T}^{r} = \int T^{r}{}_{r}r dr d\theta$, $\mathcal{T}^{\theta} = \int T^{\theta}{}_{\theta}r dr d\theta$ and $\mathcal{T}^{z} = \int T^{z}{}_{z}r dr d\theta$ (also in units of energy per unit length), to be conserved, is straightforward to show that, for the Nielsen-Olesen string
\begin{eqnarray}
\mu_{|n|} = -\mathcal{T}^{z},
\end{eqnarray}
for any value of $\beta_{|n|}$, and
\begin{eqnarray} \label{T^{rr}**}
\mathcal{T}^{r} = \mathcal{T}^{\theta} \approx -2\pi \eta^2 \left[|n|^{2\xi} - |n|^{2-2\epsilon} - |n|^2\ln\left(\sqrt{\beta_{|n|}}\right)\right].
\end{eqnarray}
Hence, for $\xi = \epsilon = 1/2$, as in Eq. (\ref{ParamBounds2}), and at critical coupling, the only nonzero component of the tension is $\mathcal{T}^{z}$, which is equal to minus the mass-energy per unit length of the string.
Substituting the ansatz Eq. (\ref{NO_ansatz}) into Eq. (\ref{Cons_current}), it is also straightforward to show that $j^{\mu} = 0, \ \forall \mu \in \left\{0,1,2,3\right\}$ and hence that the string is uncharged.
%
%%%%%%%%%%%%%%%%%%%%%%%%%%%%%%%%%%%%%%%%%%%%%%%%%%%%%%%%
%%%%%%%%%%%%%%%%%%%%%%%%%%%%%%%%%%%%%%%%%%%%%%%%%%%%%%%%``
%
\section{Field configuration and constants of motion for a lumpy string}  \label{Sect.III}
\subsection{Field configuration for a lumpy string} \label{Sect.IIIA}
In this section, we consider a model of an abelian-Higgs string, with variable scalar and vector core radii, in which the essential phase structure of the Nielsen-Olesen solution is preserved. That is, we wish to preserve the circular symmetry and topological winding number of the string cross section at all times and at each point along the string length, even if the radius of the cross section is itself a function of $t$ and $z$. We therefore take the ansatz
\begin{eqnarray} \label{scalar_ansatz}
\phi(r,\theta,z,t) = \eta F(r,z,t)e^{in\theta},
\end{eqnarray}
\begin{eqnarray} \label{vector_ansatz}
A_{\theta} = \frac{n}{e}a_{\theta}(r,z,t),  \ \ \ A_r = A_z = A_0 = 0.
\end{eqnarray}
The scalar EOM then becomes
\begin{eqnarray} \label{specific_scalar_EOM}
\frac{\partial^2 F}{\partial r^2} + \frac{1}{r} \frac{\partial F}{\partial r} - \frac{n^2 F}{r^2}(1-a_{\theta})^2 + \frac{\partial^2 F}{\partial z^2} - \frac{\partial^2 F}{\partial t^2} - \frac{1}{2 r_s^2}F(F^2-1) = 0,
\end{eqnarray}
and the $\theta$-component of the vector EOM is
\begin{eqnarray} \label{specific_vector_EOM_theta}
\frac{\partial^2 a_{\theta}}{\partial r^2} - \frac{1}{r} \frac{\partial a_{\theta}}{\partial r} +\frac{\partial^2 a_{\theta}}{\partial z^2} - \frac{\partial^2 a_{\theta}}{\partial t^2} +\frac{2F^2}{r_v^2}(1-a_{\theta}) = 0.
\end{eqnarray}
The $z$- and $t$-components both give
\begin{eqnarray} \label{specific_vector_EOM_rz}
\frac{\partial^2 a_{\theta}}{\partial z^2} - \frac{\partial^2 a_{\theta}}{\partial t^2} = 0,
\end{eqnarray}
which allows us to cancel these terms from the $\theta$-component, and the $r$-component is identically satisfied. \\
\indent
Clearly, there exist functional forms for $F(r,z,t)$ and $a_{\theta}(r,z,t)$, which satisfy the EOM, that include both left \emph{and} right movers, i.e.
\begin{eqnarray} \label{travelling_wave1}
F(r,z,t) = F(r - g_{s}(z\pm t)), \ \ \ a_{\theta}(r,z,t) = a_{\theta}(r-g_{v}(z\pm t)),
\end{eqnarray}
where $g_{s}(z\pm t)$ and $g_{v}(z\pm t)$ are arbitrary functions of $z \pm t$ and which, in principle, may include superpositions of the form $g_s(z \pm t) = g_{s(L)}(z + t) + g_{s(R)}(z - t)$. However, we must carefully examine the physical nature of these waves. In Cartesian coordinates $(t,x,y,z)$, an ansatz representing waves propagating \emph{on} the string, that is, representing displacements of the string central axis from the original coordinate $z$-axis, takes the form \cite{VaVa:90}
\begin{eqnarray} \label{travelling_wave2}
\phi(x,y,z,t) = \Phi(X,Y), \ \ \ A_{\mu}(x,y,z,t) = \mathcal{A}_{\mu}(X,Y),
\end{eqnarray}
where
\begin{eqnarray} \label{travelling_wave3}
X = x - g_x(z \pm t), \ \ \ Y = y - g_y(z \pm t),
\end{eqnarray}
and $\Phi(x,y)$, $\mathcal{A}_{\theta}(x,y)$ represent the original Nielsen-Olesen solution, expressed in Cartesians. Such a solution represents a string, of finite width, whose central axis is located at the ``origin" with
respect to the transformed coordinates $X = Y = 0$, rather than at the origin of the physical coordinates of the background space, $x = y = 0$. In this case, the general string configuration, at a given time $t$, is that of a curved cylinder, but each vortex cross section (taken perpendicular to the central axis) remains a circle of fixed radius. Solutions of this type, known as ``traveling wave" solutions, for abelian-Higgs strings, were shown to exist in \cite{VaVa:90} (see also the appendix in \cite{LaYo:12} for further discussion).
In fact, to obtain true traveling wave solutions, it is not sufficient to have only nonzero $\phi$, $A_{x}$ and $A_{y}$, but nonzero $A_{z}$ and $A_{0}$ of the form
\begin{eqnarray} \label{travelling_wave4}
A_{0} = \pm [\dot{g}_y\mathcal{A}_{x}(X,Y) + \dot{g}_x\mathcal{A}_{y}(X,Y)] = \pm A_{z},
\end{eqnarray}
are also required. In this case, though the functions $g_x(z \pm t)$ and $g_y(z \pm t)$ may represent either left \emph{or} right movers, the choice of sign in the argument of each must match that in Eq. (\ref{travelling_wave4}), so that linear superpositions of waves propagating in opposite directions are not allowed \cite{VaVa:90}. \\
\indent
As we shall see, similar restrictions apply to the left/right movers that characterize the $z$- and $t$-dependence of the lumpy string ansatz but, first, let us take a closer look at the solutions proposed in Eq. (\ref{travelling_wave1}). Since, by analogy with Eqs. (\ref{travelling_wave2})-(\ref{travelling_wave3}), the values of $g_s(z \pm t)$ and  $g_v(z \pm t)$ represent the displacement from $r$ of the variables $\mathcal{R}_s = r - g_s(z \pm t)$ and $\mathcal{R}_v = r - g_v(z \pm t)$, respectively, at the point $z$ at time $t$, we see that the conditions $\mathcal{R}_s = \mathcal{R}_v = 0$
%\begin{eqnarray} \label{travelling_wave5}
%\mathcal{R}_s = const., \ \ \ \mathcal{R}_v = const.,
%\end{eqnarray}
do not correspond to a unique point within a cross section of the string. %(i.e. within a given two-dimensional vortex).
%, unless $r=0$.
Moreover, since the true vacuum can exist within a vortex at only one point \cite{ViSh:00}, we are required to impose the boundary conditions
\begin{equation} \label{F_bc}
F(r,z,t) = \left \lbrace
\begin{array}{rl}
0,& \ r=0 \\
1,& \ r \rightarrow \infty, \ \forall z,t
\end{array}\right.
\end{equation}
\begin{equation} \label{a_{theta}_bc}
a_{\theta}(r,z,t) = \left \lbrace
\begin{array}{rl}
0,& \ r=0 \\
1,& \ r \rightarrow \infty, \ \forall z,t,
\end{array}\right.
\end{equation}
which is impossible, for the solutions proposed in Eq. (\ref{travelling_wave1}), without setting $g_s = g_v = 0, \ \forall z,t$. Such solutions, though they exist mathematically, therefore violate the physical requirement that $\langle \phi \rangle = 0$ at one point \emph{only} within a given vortex \cite{ViSh:00}.\\
\indent
As an alternative to Eq. (\ref{travelling_wave1}), we may instead search for approximate solutions of the form
\begin{equation} \label{FaNew}
F(r,z,t) = F\left(R_{s|n|}^{eff}\right), \ \ \ a_{\theta}(r,z,t) = a_{\theta}\left(R_{v|n|}^{eff}\right),
\end{equation}
where we have defined
\begin{equation} \label{R^{eff}}
R_{i|n|}^{eff}(r,z,t) = \frac{r}{r_{i|n|}^{eff}(z,t)}, \ i \in \left\{s,v\right\},
\end{equation}
and where $r_{s|n|}^{eff}(z,t)$ and $ r_{v|n|}^{eff}(z,t)$ represent the effective \emph{local} values of the scalar and vector core radii (for given values of $z$ and $t$), respectively.  The physical nature of the waves described by the ansatz, Eqs. (\ref{scalar_ansatz})-(\ref{vector_ansatz}), now becomes clear %While the traveling wave solutions presented in \cite{VaVa:90} represent waves on strings that remain cylindrical with respect to an oscillating axis, the ansatz proposed here describes waves on strings whose axis remains \emph{fixed} with respect to the background. These waves result from the violation of cylindrical symmetry caused by $t$- and $z$-dependence in the effective radii of the string core. \\
%\indent
and we may write $r_{s|n|}^{eff}(z,t)$ and $r_{v|n|}^{eff}(z,t)$ in the form
\begin{eqnarray} \label{r_in^eff}
r_{i|n|}^{eff}(z,t) = r_{i|n|} g_i(z,t), \ \ \ i \in \left\{s,v\right\},
\end{eqnarray}
where the $g_i(z,t)$ are again arbitrary (though this time, dimensionless) functions. Substituting for $F(r,z,t)$ and $a_{\theta}(r,z,t)$ from Eq. (\ref{FaNew}) into Eqs. (\ref{specific_scalar_EOM})-(\ref{specific_vector_EOM_rz}), the scalar EOM becomes
\begin{eqnarray}
\frac{d^2 F}{d (R_{s|n|}^{eff})^2} &+& \frac{1}{R_{s|n|}^{eff}} \frac{d F}{d R_{s|n|}^{eff}} - \frac{n^2 F}{(R_{s|n|}^{eff})^2}(1-a_{\theta})^2 - \frac{1}{2}\left(\frac{r_{s|n|}^{eff}}{r_s}\right)^2F(F^2-1)
\nonumber\\
&+& (R_{s|n|}^{eff})^2 \frac{d^2 F}{d (R_{s|n|}^{eff})^2}\left[\left(\frac{\partial r_{s|n|}^{eff}}{\partial z}\right)^2 - \left(\frac{\partial r_{s|n|}^{eff}}{\partial t}\right)^2\right]
\nonumber\\
&-& R_{s|n|}^{eff}\frac{d F}{d R_{s|n|}^{eff}}\left\{r_{s|n|}^{eff}\left(\frac{\partial^2 r_{s|n|}^{eff}}{\partial z^2} - \frac{\partial^2 r_{s|n|}^{eff}}{\partial t^2}\right) - 2\left[\left(\frac{\partial r_{s|n|}^{eff}}{\partial z}\right)^2 - \left(\frac{\partial r_{s|n|}^{eff}}{\partial t}\right)^2\right] \right\} = 0,
\nonumber
\end{eqnarray}
and the $z$- and $t$- components of the vector EOM are equivalent to
\begin{eqnarray}
&{}& (R_{v|n|}^{eff})^2 \frac{d^2 a_{\theta}}{d (R_{v|n|}^{eff})^2}\left[\left(\frac{\partial r_{v|n|}^{eff}}{\partial z}\right)^2 - \left(\frac{\partial r_{v|n|}^{eff}}{\partial t}\right)^2\right]
\nonumber\\
&-& R_{v|n|}^{eff}\frac{d a_{\theta}}{d R_{v|n|}^{eff}}\left\{r_{v|n|}^{eff}\left(\frac{\partial^2 r_{v|n|}^{eff}}{\partial z^2}-\frac{\partial^2 r_{v|n|}^{eff}}{\partial t^2}\right) - 2\left[\left(\frac{\partial r_{v|n|}^{eff}}{\partial z}\right)^2 -\left(\frac{\partial r_{v|n|}^{eff}}{\partial t}\right)^2\right] \right\} = 0,
\nonumber
\end{eqnarray}
while the $\theta$-component takes a form analogous to Eq. (\ref{NO_specific_vector_EOM_theta}), under the identifications $\alpha_{\theta} \leftrightarrow a_{\theta}$, $f \leftrightarrow F$, $r_{v|n|}^{eff}(z,t) \leftrightarrow r_{v|n|}$ and $R_{v|n|}^{eff}(z,t) \leftrightarrow R_{v|n|}$.\\
\indent
The EOM therefore reduce to
\begin{eqnarray} \label{specific_scalar_EOM*}
\frac{d^2 F}{d (R_{s|n|}^{eff})^2} &+& \frac{1}{R_{s|n|}^{eff}} \frac{d F}{d R_{s|n|}^{eff}} - \frac{n^2 F}{(R_{s|n|}^{eff})^2}(1-a_{\theta})^2 - \frac{1}{2}\left(\frac{r_{s|n|}^{eff}}{r_s}\right)^2F(F^2-1) = 0,
\end{eqnarray}
\begin{eqnarray} \label{specific_vector_EOM_theta*}
\frac{d^2 a_{\theta}}{d (R_{v|n|}^{eff})^2} - \frac{1}{R_{v|n|}^{eff}} \frac{d a_{\theta}}{d R_{v|n|}^{eff}} + 2\left(\frac{r_{v|n|}^{eff}}{r_v}\right)^2 F^2(1-a_{\theta}) = 0,
\end{eqnarray}
if the functions $g_i(z,t)$ satisfy the conditions
\begin{eqnarray}  \label{EOM_g1}
\left(\frac{\partial g_i}{\partial z}\right)^2 = \left(\frac{\partial g_i}{\partial t}\right)^2, \ \ \  \frac{\partial g_i}{\partial z} = \pm \frac{\partial g_i}{\partial t},
\end{eqnarray}
which, in turn, automatically imply
\begin{eqnarray}  \label{EOM_g2}
\frac{\partial^2 g_i}{\partial z^2} = \frac{\partial^2 g_i}{\partial t^2}.
\end{eqnarray}
However, Eq. (\ref{EOM_g1}) is the stronger condition, since Eq. (\ref{EOM_g2}) is satisfied by superpositions of arbitrary left \emph{and} right movers (i.e. functions of the form $g_i(z,t) = g_{i(L)}(z + t) + g_{i(R)}(z - t)$), whereas Eq. (\ref{EOM_g2}) is satisfied only by arbitrary left \emph{or} right movers: $g_i(z,t) = g_{i}(z + t)$ \emph{or}  $g_i(z,t) = g_{i}(z - t)$. \\
\indent
Clearly, by analogy with the solutions obtained for Nielsen-Olesen strings, Eqs.  (\ref{specific_scalar_EOM*})-(\ref{specific_vector_EOM_theta*}) are solved, to leading order in the uncoupled regime, by
\begin{equation} \label{F(r,z,t)}
F(r,z,t) \approx \left \lbrace
\begin{array}{rl}
\left(r/r_{s|n|}^{eff}(z,t)\right)^{|n|},& \ r \lesssim r_{s|n|}^{eff}(z,t) \\
1,& \ r \gtrsim r_{s|n|}^{eff}(z,t),
\end{array}\right.
\end{equation}
\begin{equation} \label{a_{theta}(r,z,t)}
a_{\theta}(r,z,t) \approx \left \lbrace
\begin{array}{rl}
\left(r/r_{v|n|}^{eff}(z,t)\right)^{2},& \ r \lesssim r_{v|n|}^{eff}(z,t) \\
1,& \ r \gtrsim r_{v|n|}^{eff}(z,t).
\end{array}\right.
\end{equation}
Physically, these functions represent the spatial and temporal variation of the scalar and vector core radii, for a string with a \emph{fixed} central axis (relative to the background space), lying parallel to the $z$-axis. Combining the conditions in Eqs. (\ref{EOM_g1})-(\ref{EOM_g2}) with the solutions given in Eqs. (\ref{F(r,z,t)})-(\ref{a_{theta}(r,z,t)}) automatically implies
\begin{eqnarray} \label{WaveEq*_Scalar}
\left(\frac{\partial F}{\partial z}\right)^2 = \left(\frac{\partial F}{\partial t}\right)^2, \ \ \  \frac{\partial F}{\partial z} = \pm \frac{\partial F}{\partial t},
\end{eqnarray}
\begin{eqnarray} \label{WaveEq*_Vector}
\left(\frac{\partial a_{\theta}}{\partial z}\right)^2 = \left(\frac{\partial a_{\theta}}{\partial t}\right)^2, \ \ \  \frac{\partial a_{\theta}}{\partial z} = \pm \frac{\partial a_{\theta}}{\partial t},
\end{eqnarray}
and, though it follows that
\begin{eqnarray}  \label{WaveEq_Scalar}
\frac{\partial^2 F}{\partial z^2} = \frac{\partial^2 F}{\partial t^2},
\end{eqnarray}
\begin{eqnarray}  \label{WaveEq_Vector}
\frac{\partial^2 a_{\theta}}{\partial z^2} = \frac{\partial^2 a_{\theta}}{\partial t^2},
\end{eqnarray}
Eqs. (\ref{WaveEq*_Scalar})-(\ref{WaveEq*_Vector}) represent stronger constraints, which prevent the possibility of superpositions of left and right-movers in the radial oscillations, as stated above.\\
\indent
Finally, before moving on to calculate the energy and conserved momentum of the lumpy string in the next section, we note two further restrictions that physical considerations place on the mathematical form of the functions $g_i(z,t)$. Firstly, since we have used an ansatz describing (formally) infinite strings, we must impose periodic boundary conditions in the $z$-coordinate and consider a finite section of length $\Delta$ when determining the constants of motion, in order to prevent the energy and momentum from diverging for $z \rightarrow \pm \infty$. Similarly, the fact that both quantities must be finite and conserved, for string sections of finite length, implies periodicity in $t$ in order to prevent divergence for $t \rightarrow \infty$. We therefore impose the conditions
\begin{equation} \label{Periodic_BC}
g_i((z + m\lambda_z) \pm (t + pT)) = g_i(z \pm t), \ \ \ m,p \in \mathbb{Z}
\end{equation}
for some arbitrary length scale $\lambda_z$ and time period $T$, which may be associated with a frequency via
\begin{equation} \label{omega_z}
\omega_z = \frac{2\pi}{T} = \pm k_z = \pm \frac{2\pi}{\lambda_z}.
\end{equation}
Secondly, the finiteness of the energy density also requires us to choose the form of $g_i(z \pm t)$ such that
\begin{equation} \label{}
g_i(z \pm t) \neq 0, \ \ \ \forall z,t
\end{equation}
as we shall see explicitly in Sect. \ref{Sect.IIIB}. In the classical theory therefore, an arbitrary cut-off must be imposed by hand but, thanks to the well known quantum properties of abelian-Higgs strings, we may simply set
\begin{equation} \label{}
g_i(z \pm t) \geq 1, \ \ \ \forall z,t
\end{equation}
which ensures that $r_{i|n|}^{eff}(z,t) \geq r_{i|n|}, \ \forall z,t$. %according to the definition in Eq. (\ref{r_in^eff}). 
This ensures that the local radii of the scalar and vector string cores do not drop below the minimum values dictated by the Compton wavelengths of the associated bosons. For the specific choice of $g_s(z \pm t) = g_v(z \pm t) = 1$, we recover the Nielsen-Olesen solution \cite{NiOl:73}.
%$g_i(z,t)$, $i \in \left\{s,v\right\}$, $t \rightarrow \infty$.
%
%%%%%%%%%%%%%%%%%%%%%%%%%%%%%%%%%%%%%%%%%%%%%%%%%%%%%%%%
%%%%%%%%%%%%%%%%%%%%%%%%%%%%%%%%%%%%%%%%%%%%%%%%%%%%%%%%``
\subsection{Lumpy string energy and momentum}  \label{Sect.IIIB}
If $k^{(\alpha)}{}_{\nu}$, $\alpha = \left\{0,1,2, . . .d\right\}$, denotes a set of $D = d+1$ Killing vectors, then
\begin{eqnarray}
j^{(\alpha)\mu} = k^{(\alpha)}{}_{\nu}T^{\mu\nu},
\end{eqnarray}
denotes a set of $D$ conserved currents,
\begin{eqnarray}
\partial_{\mu}j^{(\alpha)\mu} = 0,
\end{eqnarray}
and the associated conserved charges are
\begin{eqnarray}
\Pi^{\alpha} = \int j^{(\alpha)0}\sqrt{-g}d^3x = \int k^{(\alpha)}{}_{\nu}T^{0\nu}\sqrt{-g}d^3x.
\end{eqnarray}
For the sake of completeness and clarity, we list all independent components of the energy-momentum tensor below:
\begin{eqnarray} %\label{T^{00}}
T^{00} = \eta^2\left\{2\left(\frac{\partial F}{\partial t}\right)^2 + \frac{|n|^2r_v^2}{r^2}\left(\frac{\partial a_{\theta}}{\partial t}\right)^2 + \left(\frac{\partial F}{\partial r}\right)^2 + \frac{|n|^2F^2}{r^2}(1-a_{\theta})^2 + \frac{|n|^2r_v^2}{4r^2}\left(\frac{\partial a_{\theta}}{\partial r}\right)^2 + \frac{1}{4r_s^2}(F^2-1)^2\right\},
\nonumber
\end{eqnarray}
\begin{eqnarray} \label{T^{00}}
{}
\end{eqnarray}
\begin{eqnarray} \label{T^{0r}}
T^{0r} = -\eta^2\left\{2\frac{\partial F}{\partial r}\frac{\partial F}{\partial t} + \frac{|n|^2r_v^2}{r^2}\frac{\partial a_{\theta}}{\partial r}\frac{\partial a_{\theta}}{\partial t}\right\},
\end{eqnarray}
\begin{eqnarray} \label{T^{0theta}}
T^{0\theta} = 0,
\end{eqnarray}
\begin{eqnarray} \label{T^{0z}}
T^{0z} = -\eta^2\left\{2\frac{\partial F}{\partial z}\frac{\partial F}{\partial t} + \frac{|n|^2r_v^2}{r^2}\frac{\partial a_{\theta}}{\partial z}\frac{\partial a_{\theta}}{\partial t}\right\},
\end{eqnarray}
\begin{eqnarray} \label{T^{rr}}
T^{rr} = \eta^2\left\{\left(\frac{\partial F}{\partial r}\right)^2 - \frac{n^2F^2}{r^2}(1-a_{\theta})^2 + \frac{3}{4}\frac{|n|^2r_v^2}{r^2}\left(\frac{\partial a_{\theta}}{\partial r}\right)^2 -  \frac{1}{4r_s^2}(F^2-1)^2\right\},
\end{eqnarray}
\begin{eqnarray} \label{T^{rtheta}}
T^{r\theta} = 0,
\end{eqnarray}
\begin{eqnarray} \label{T^{rz}}
T^{rz} = \eta^2\left\{2\frac{\partial F}{\partial r}\frac{\partial F}{\partial z} + \frac{|n|^2r_v^2}{r^2}\frac{\partial a_{\theta}}{\partial r}\frac{\partial a_{\theta}}{\partial z}\right\},
\end{eqnarray}
\begin{eqnarray} \label{T^{thetatheta}}
T^{\theta\theta} = \eta^2\left\{-\frac{1}{r^2}\left(\frac{\partial F}{\partial r}\right)^2 + \frac{|n|^2F^2}{r^4}(1-a_{\theta})^2 + \frac{3}{4}\frac{|n|^2r_v^2}{r^4}\left(\frac{\partial a_{\theta}}{\partial r}\right)^2 -  \frac{1}{4r_s^2 r^2}(F^2-1)^2\right\},
\end{eqnarray}
\begin{eqnarray} \label{T^{thetaz}}
T^{\theta z} = 0,
\end{eqnarray}
\begin{eqnarray}% \label{T^{zz}}
T^{zz} =  \eta^2\ \left\{2\left(\frac{\partial F}{\partial z}\right)^2 + \frac{|n|^2r_v^2}{r^2}\left(\frac{\partial a_{\theta}}{\partial z}\right)^2 - \left(\frac{\partial F}{\partial r}\right)^2 - \frac{|n|^2F^2}{r^2}(1-a_{\theta})^2 - \frac{|n|^2r_v^2}{4r^2}\left(\frac{\partial a_{\theta}}{\partial r}\right)^2 -  \frac{1}{4r_s^2}(F^2-1)^2\right\}.
\nonumber
\end{eqnarray}
\begin{eqnarray} \label{T^{zz}}
{}
\end{eqnarray}
For $(3+1)$-dimensional Minkowski space there are ten Killing vectors, namely
%\begin{eqnarray}
%k^{(\alpha)}{}_{\nu} &=& \delta^{\alpha}{}_{\nu}, \ \alpha \in \left\{0,1,2,3\right\},
%\nonumber \\
%\kappa^{(l)}{}_{\nu} &=& (\kappa^{(l)}{}_{0},\kappa^{(l)}{}_{m}) = (0,\epsilon^{l}{}_{mn}x^{n}), \ l,m,n \in  \left\{1,2,3\right\},
%\nonumber \\
%K^{(l)}{}_{\nu} &=& (K^{(l)}{}_{0},K^{(l)}{}_{m}) = (x^{l},-\delta^{l}{}_{m}x^{0}),
%\end{eqnarray}
%where $x^{l}$, $x^{m}$ and $x^{n}$ denote unit vectors parallel to Cartesian axes, and where we have used the letters $k$, $\kappa$ and $K$ to distinguish between vectors
four related to space-time translational invariance, three to spatial rotational invariance, and three to boost invariance, respectively. However,
from Eqs. (\ref{T^{00}})-(\ref{T^{zz}}), it can be seen that the only relevant independent Killing vectors, giving rise to nonzero conserved quantities for the ``straight" lumpy string described by Eqs. (\ref{scalar_ansatz})-(\ref{vector_ansatz}), are
\begin{eqnarray}
k^{(0)}{}_{\nu} = (1,0,0,0), \ \ \ k^{(z)}{}_{\nu} = (0,0,0,1).% \ \ \ K^{(z)}{}_{\nu} = (z,0,0,-t).
\end{eqnarray}
Thus, the independent conserved quantities are the Hamiltonian $E_{|n|}$ and the momentum in the $z$-direction, $P^{z} = -P_{z}$. The total $3$-momentum is $P_{|n|} = \sqrt{-P^{i}P_{i}} = \sqrt{-P^{z}P_{z}}$, and the conserved $4$-momentum is given by
\begin{eqnarray}
\Pi_{|n|} = \sqrt{E_{|n|}^2 - P_{|n|}^2}.
\end{eqnarray}
We may also define the integrated tensions (negative pressures) as
\begin{eqnarray} \label{Pressures1}
T^{i} = \int T^{i}{}_{(i)} \sqrt{-g}d^3x,
\end{eqnarray}
where the notation $T^{i}{}_{(i)}$ indicates that no sum is implied by the repeated index, and the integrated shears between two arbitrary directions, $x^{i}$ and $x^{j}$, via
\begin{eqnarray} \label{Shears1}
S^{ij} = \int T^{ij} \sqrt{-g}d^3x,
\end{eqnarray}
for $i,j \in \left\{r,\theta,z\right\}$. \\
\indent
Using the approximate solutions for $F(r,z,t)$ and $a_{\theta}(r,z,t)$, Eqs. (\ref{F(r,z,t)})-(\ref{a_{theta}(r,z,t)}), the order of magnitude values for the nontrivial constants of motion, integrated pressures and shears for a finite section of lumpy string, of length $\Delta$, are
\begin{eqnarray} \label{T^{00}*}
E_{|n|} &\approx& 2\pi \eta^2 \int_{0}^{\Delta} dz \left\{|n| +|n| r_{s|n|}^2\left[\frac{g_s^2}{r_s^2} + \left(\frac{\partial g_s}{\partial t}\right)^2\right] + |n|^2r_{v}^2\left[\frac{1}{r_{v|n|}^2g_v^2} + \frac{1}{g_v^2}\left(\frac{\partial g_v}{\partial t}\right)^2\right] \right\}%+  |n|^2\ln\left(\sqrt{\beta_{|n|}^{eff}}\right)\right\}
\nonumber \\
%&{}&
&+&  2\pi \eta^2|n|^2 \int_{0}^{\Delta} dz \ln\left(\sqrt{\beta_{|n|}^{eff}}\right),
\end{eqnarray}
\begin{eqnarray} \label{T^{0z}*}
P^{z}  \approx -2\pi\eta^2 \int_{0}^{\Delta} dz \left\{|n| r_{s|n|}^2\frac{\partial g_s}{\partial z}\frac{\partial g_s}{\partial t} + \frac{|n|^2r_v^2}{g_v^2}\frac{\partial g_v}{\partial z}\frac{\partial g_v}{\partial t}\right\},
\end{eqnarray}
\begin{eqnarray} \label{T^{rr}*}
T^{r} = T^{\theta} \approx -2\pi \eta^2 \int_{0}^{\Delta} dz \left\{\frac{r_{s|n|}^2g_s^2}{r_s^2} - \frac{|n|^2r_v^2}{r_{v|n|}^2g_v^2} - |n|^2\ln\left(\sqrt{\beta_{|n|}^{eff}}\right)\right\},
\end{eqnarray}
%
%\begin{eqnarray} \label{T^{thetatheta}*}
%\mathcal{T}^{\theta} \approx 2\pi \eta^2  \int_{0}^{\Delta} dz \left\{\frac{r_{s|n|}^2g_s^2}{r_s^2} - \frac{|n|^2 r_v^2}{r_{v|n|}^2g_v^2} - |n|^2\ln\left(\sqrt{\beta_{|n|}^{eff}}\right)\right\},
%\end{eqnarray}
%
\begin{eqnarray} \label{T^{zz}*}
T^{z} &\approx&  -2\pi\eta^2  \int_{0}^{\Delta} dz \left\{|n| + |n|r_{s|n|}^2\left[\frac{g_s^2}{r_s^2} - \left(\frac{\partial g_s}{\partial z}\right)^2\right] + |n|^2r_{v}^2\left[\frac{1}{r_{v|n|}^2g_v^2} - \frac{1}{ g_v^2}\left(\frac{\partial g_v}{\partial z}\right)^2\right]\right\}% +  |n|^2\ln\left(\sqrt{\beta_{|n|}^{eff}}\right)\right\}
\nonumber \\
%&{}&
&+&  2\pi \eta^2|n|^2 \int dz \ln\left(\sqrt{\beta_{|n|}^{eff}}\right),
\end{eqnarray}
\begin{eqnarray} \label{T^{rz}*}
S^{rz} = -2\pi \eta^2  \int_{0}^{\Delta} dz \left\{|n|r_{s|n|} \frac{\partial g_s}{\partial z} + \frac{|n|^2 r_v^2}{r_{v|n|}g_v^3} \frac{\partial g_v}{\partial z}\right\} = 0,
\end{eqnarray}
where we have used the approximation $|n|+1 \approx |n|$ for $|n| \geq 1$, and ignored numerical factors of order unity multiplying individual terms within the integrands. The vanishing of the integrated shear $S^{rz}$ follows directly from the imposition of periodic boundary conditions on $g_s(z,t)$ and $g_v(z,t)$, Eq. (\ref{Periodic_BC}), if we choose to set
\begin{eqnarray} \label{Delta}
\Delta = m\lambda_z,
\end{eqnarray}
though a nonzero local shear, $\mathcal{S}^{rz}(z,t) = \int T^{rz}(z,t) rdr$, is, in general, present.\\
\indent
It is straightforward to check that the standard expressions for the constants of motion and integrated pressures/shears for the Nielsen-Olesen string are recovered for $g_s(z,t) = g_c(z,t) = 1$. However, Eqs. (\ref{T^{00}*})-(\ref{T^{rz}*}) also simplify considerably under the more general assumption
\begin{eqnarray}
g_s(z,t) = g_v(z,t) = g_c(z,t),
\end{eqnarray}
together with critical coupling and $\xi = \epsilon$. In this case,
\begin{eqnarray}
r_{s|n|}^{eff}(z,t) = r_{v|n|}^{eff}(z,t) = r_{c|n|}^{eff}(z,t), \ \ \ (\beta_{|n|}^{eff}(z,t) = 1),
\end{eqnarray}
and the logarithmic terms vanish, while those involving $g_s$, $g_v$, $r_{s|n|}$ and $r_{v|n|}$ can be consolidated. For the specific choice, $\xi = \epsilon = 1/2$ (as before), and using the fact that $\partial g_c/\partial t = \pm \partial g_c/\partial z$, we then have
\begin{eqnarray} \label{T^{00}**}
E_{|n|} \approx 2\pi \eta^2|n|\Delta \left[1 + \frac{1}{\Delta}\int_{0}^{\Delta} dz \left\{|n|g_c^2 + \frac{1}{g_c^2} + r_{c|n|}^2\left(\frac{\partial g_c}{\partial z}\right)^2\left[1 + \frac{1}{g_c^2}\right] \right\}\right],
\end{eqnarray}
\begin{eqnarray} \label{T^{0z}**}
P^{z}  \approx \pm 2\pi\eta^2 |n| r_{c|n|}^2 \int_{0}^{\Delta} dz \left\{\left(\frac{\partial g_c}{\partial z}\right)^2 + \frac{1}{g_c^2}\left(\frac{\partial g_c}{\partial z}\right)^2\right\},
\end{eqnarray}
\begin{eqnarray} \label{T^{rr}**}
T^{r} = T^{\theta} \approx -2\pi \eta^2|n| \int_{0}^{\Delta} dz \left\{g_c^2 - \frac{1}{g_c^2}\right\},
\end{eqnarray}
\begin{eqnarray} \label{T^{zz}**}
T^{z} &\approx&  2\pi \eta^2|n|\Delta \left[1 + \frac{1}{\Delta}\int_{0}^{\Delta} dz \left\{|n|g_c^2 + \frac{1}{g_c^2} - r_{c|n|}^2\left(\frac{\partial g_c}{\partial z}\right)^2\left[1 + \frac{1}{g_c^2}\right] \right\}\right],
\end{eqnarray}
where the choice of sign in Eq. (\ref{T^{0z}**}) is positive for right-movers and negative for left-movers.
%
%\begin{eqnarray} \label{T^{rz}**}
%S^{rz} = -2\pi \eta^2|n|r_{c|n|} \int_{0}^{\Delta} dz \left\{\frac{\partial g_c}{\partial z} + \frac{1}{g_c^3} \frac{\partial g_c}{\partial z}\right\} = 0.
%\end{eqnarray}
Given also that $g_{c}(z,t) \geq 1$, we may, to lowest order in the approximation, neglect terms in $g_c^{-2}$ and $g_c^{-2}(\partial g_c/\partial z)^2$, as long as their signs do not conflict with those of terms proportional to $g_c^{2}$ and $g_c^{2}(\partial g_c/\partial z)^2$, respectively. \\
\indent
Furthermore, by rewriting the $z$ coordinate in terms of a dimensionless parameter $\sigma \in [0,2\pi)$, such that
\begin{eqnarray}
z(\sigma) = (2\pi)^{-1}\Delta \sigma,
\end{eqnarray}
we may write
\begin{eqnarray}\label{z-sigma}
g_c(z \pm t) = g_c(k_z z + \omega_z t) = g_c(m\sigma + \omega_{z} t),
\end{eqnarray}
where $\omega_z = \pm k_z = \pm 2\pi/\lambda_z = \pm 2\pi m/\Delta$, in accordance with Eqs. (\ref{omega_z}) and (\ref{Delta}). Of course, different plane wave modes in the Fourier expansion of $g_c(z \pm t)$ correspond to different frequencies $\omega_j$ and associated wavelengths $\lambda_j$, such that $\omega_j = \pm k_j = \pm 2\pi/\lambda_j = \pm 2\pi m_j/ \Delta$, for some $m_j \in \mathbb{N}$. By writing $g_c(z \pm t)$ in the form given in Eq. (\ref{z-sigma}), which implies $\partial g_c/\partial t = \omega_z dg_c/du$,  $\partial g_c/\partial z = k_z dg_c/du$,  $\partial g_c/\partial \sigma = m dg_c/du$, where $u = k_z z + \omega_z t = m\sigma + \omega_z t$, we select a \emph{single} mode as being ``characteristic" of the wave form. The most natural choice is the mode with the highest amplitude, whose wavelength will give the approximate width of any ``lump" on the string. The integer $m$ may then be interpreted as defining the number of lumps present in a string section of length $\Delta$. This gives
\begin{eqnarray} \label{T^{00}***}
E_{|n|} \approx 2\pi \eta^2|n|\Delta \left[1 + \frac{1}{2\pi}\int_{0}^{2\pi} \left\{|n|g_c^2 + \frac{(2\pi)^2r_{c|n|}^2}{\Delta^2} (\partial_{\sigma}g_c)^2 \right\}d\sigma \right],
\end{eqnarray}
\begin{eqnarray} \label{T^{0z}***}
P^{z}  \approx \pm 2\pi\eta^2|n|\Delta \times \frac{(2\pi)^2r_{c|n|}^2}{\Delta^2} \times \frac{1}{2\pi}\int_{0}^{2\pi} (\partial_{\sigma}g_c)^2 d\sigma,
\end{eqnarray}
\begin{eqnarray} \label{T^{rr}***}
T^{r} = T^{\theta} \approx -2\pi \eta^2|n| \Delta \times \frac{1}{2\pi}\int_{0}^{2\pi} \left\{g_c^2 - \frac{1}{g_c^2}\right\} d\sigma,
\end{eqnarray}
\begin{eqnarray} \label{T^{zz}***}
T^{z} \approx  -2\pi \eta^2|n|\Delta \left[1 + \frac{1}{2\pi}\int_{0}^{2\pi} \left\{|n|g_c^2 - \frac{(2\pi)^2r_{c|n|}^2}{\Delta^2} (\partial_{\sigma}g_c)^2 \right\}d\sigma \right].
\end{eqnarray}
We may then write
\begin{eqnarray}
\frac{1}{2\pi}\int_{0}^{2\pi} g_c^2 d\sigma = \alpha^2, \ \ \ \frac{1}{2\pi}\int_{0}^{2\pi} (\partial_{\sigma}g_c)^2 d\sigma = m^2\gamma^2,  \ \ \ \frac{1}{2\pi}\int_{0}^{2\pi} g_c^{-2} d\sigma = \delta^2,
\end{eqnarray}
where $\alpha^2 \geq 1$, $\gamma^2 \geq 0$, $\delta^2 \leq 1$ are constants, with equality holding if and only if $g_c(z,t) = 1$, $\forall z,t$, so that
\begin{eqnarray} \label{T^{00}****}
E_{|n|} \approx 2\pi \eta^2|n|\Delta \left[1 +  |n|\alpha^2 + \frac{(2\pi)^2r_{c|n|}^2\gamma^2}{\lambda_z^2} \right],
\end{eqnarray}
\begin{eqnarray} \label{T^{0z}****}
P^{z}  \approx \pm 2\pi\eta^2|n|\Delta \times \frac{(2\pi)^2r_{c|n|}^2 \gamma^2}{\lambda_z^2},
\end{eqnarray}
\begin{eqnarray} \label{T^{rr}****}
T^{r} = T^{\theta} \approx -2\pi \eta^2|n| \Delta (\alpha^2 - \delta^2),
\end{eqnarray}
\begin{eqnarray} \label{T^{zz}****}
T^{z} \approx  -2\pi \eta^2|n|\Delta \left[1 +  |n|\alpha^2 - \frac{(2\pi)^2r_{c|n|}^2\gamma^2}{\lambda_z^2} \right],
\end{eqnarray}
and
\begin{eqnarray}
\Pi_{|n|} \approx 2\pi\eta^2|n|\Delta(1+|n|\alpha^2)\sqrt{1 +  \frac{(2\pi)^2r_{c|n|}^2}{\lambda_z^2}\left(\frac{2\gamma^2}{1+|n|\alpha^2}\right)}.
\end{eqnarray}
Interestingly, it is possible for the integrated pressure in the $z$-direction to be zero, even in the nonlinear case, if
\begin{eqnarray}
\lambda_z^2 =  (2\pi)^2r_{c|n|}^2\left(\frac{\gamma^2}{1 +  |n|\alpha^2}\right),
\end{eqnarray}
but $T^{r} = T^{\theta}$ can only vanish for the Nielsen-Olesen string. However, in general, there are local/instantaneous momentum densities, tensions and shears, which depend on both $\sigma \propto z$ and $t$, i.e.
\begin{eqnarray} \label{T^{00}+}
\mu_{|n|}(u) \approx 2\pi \eta^2|n|\left[1 + |n|g_c^2(u) + \frac{(2\pi)^2r_{c|n|}^2}{\lambda_z^2} \left(\frac{dg_c}{du}\right)^2 \right],
\end{eqnarray}
\begin{eqnarray} \label{T^{0z}+}
\mathcal{P}^{z}(u)  \approx \pm 2\pi\eta^2|n| \times \frac{(2\pi)^2r_{c|n|}^2}{\lambda_z^2}\left(\frac{dg_c}{du}\right)^2,
\end{eqnarray}
\begin{eqnarray} \label{T^{rr}+}
\mathcal{T}^{r}(u) = \mathcal{T}^{\theta}(u) \approx -2\pi \eta^2|n| \left[g_c^2(u) - \frac{1}{g_c^2(u)}\right],
\end{eqnarray}
\begin{eqnarray} \label{T^{zz}+}
\mathcal{T}^{z}(u) \approx  -2\pi \eta^2|n|\left[1 + |n|g_c^2(u) - \frac{(2\pi)^2r_{c|n|}^2}{\lambda_z^2} \left(\frac{dg_c}{du}\right)^2 \right],
\end{eqnarray}
\begin{eqnarray} \label{T^{rz}+}
\mathcal{S}^{rz}(u) \approx -(2\pi)^2 \eta^2|n| r_{c|n|} \frac{dg_c}{du},
\end{eqnarray}
where
\begin{eqnarray}
u(t,\sigma) = m\sigma + \omega_{z}t.
\end{eqnarray}
In general, therefore, we find that
\begin{eqnarray}
\mu_{|n|}(u) \geq |\mathcal{T}^{z}(u)|,
\end{eqnarray}
where equality holds only in the Nielsen-Olesen case, for which both the tension and the mass-energy per unit length become constant. From the $\mu=0$ component of the vector EOM, it is also straightforward to see that, for the lumpy string, $j^{0}=0$, and the string is uncharged.
%
%%%%%%%%%%%%%%%%%%%%%%%%%%%%%%%%%%%%%%%%%%%%%%%%%%%%%%%%
%%%%%%%%%%%%%%%%%%%%%%%%%%%%%%%%%%%%%%%%%%%%%%%%%%%%%%%%``
\section{Cosmological and astrophysical implications of lumpy strings}  \label{Sect.IV}
In this section, we consider some possible astrophysical and cosmological implications of lumpy cosmic strings. As a first step, we give a simple estimate of the energy and the momentum emission for the lumpy string configuration, as it relaxes to the standard Nielsen-Olesen form.
%
%%%%%%%%%%%%%%%%%%%%%%%%%%%%%%%%%%%%%%%%%%%%%%%%%%%%%%%%``
\subsection{Energy and momentum loss of lumpy cosmic strings} \label{Sect.IVA}
Under the simplifying assumptions, $g_s=g_v=g_c$, $\lambda = e^2$, $\xi = \epsilon = 1/2$, made in the previous section, it is straightforward to estimate the loss in energy $\Delta E_{|n|}$ and momentum $\Delta P_{|n|}$, over a string section of length $\Delta$, as the string relaxes to a Nielsen-Olesen configuration. In terms of the new model parameters $\alpha^2$, $\gamma^2$ and $\lambda_z^2$, these are
\begin{eqnarray} \label{dE}
\Delta E_{|n|} \approx 2\pi\eta^2|n|^2\Delta \left[\alpha^2 + \frac{(2\pi)^2r_{c}^2\gamma^2}{\lambda_z^2} \right] \times \left(\frac{G}{c^4}\right)^{-1},
\end{eqnarray}
\begin{eqnarray} \label{dP}
\Delta P_{|n|} \approx  \frac{2\pi\eta^2|n|^2\Delta}{c} \times \frac{(2\pi)^2r_{c}^2\gamma^2}{\lambda_z^2} \times \left(\frac{G}{c^4}\right)^{-1},
\end{eqnarray}
where $\eta$ is dimensionless, $(G/c^4)^{-1} = m_Pc^2/l_P$ and, from now on, we write fundamental constants such as $G$, $c$ and $\hbar$ explicitly. An estimate for the characteristic time scale for the loss may be obtained from the characteristic frequency,
\begin{eqnarray} \label{dt}
\Delta t \approx T = \frac{2\pi}{\omega_z} = \frac{2\pi}{k_z} = \frac{\lambda_z}{c}.
\end{eqnarray}
\indent
Roughly speaking, $\alpha^2$ characterizes the additional rest mass-energy of the lumpy string (over and above that usually contained in a length $\Delta$ of Nielsen-Olesen string), while $\gamma^2 r_c^2/\lambda_z^2$ characterizes the additional kinetic energy due to the propagation of the lumps. The latter is higher when the average distance between lumps, $\lambda_z$, is small, and the average square of the gradient, $\gamma^2$ is large. In other words, when there are many, closely spaced, sharp peaks in the radial profile. Alternatively, when the lumps are widely separated but highly localized, corresponding to a sharp peak in $g_c$ over a flat background, it would be more accurate to introduce a two-scale model, in which the characteristic inter-lump distance and the average lump width appear. \\
\indent
Nonetheless, in our current one-scale model, the difference between these two scenarios simply corresponds to a smaller or a larger value of $\gamma^2$, respectively, for a given value of $\lambda_z$. Therefore, for any model in which the lumps are characterized by sharp peaks, it seems reasonable to assume that the ratio $\gamma^2r_c^2/\lambda_z^2$ is large, even if we take $\lambda_z \gtrsim r_c$ as a fundamental cut-off due to quantum effects, and that the kinetic term is of leading order. For radial profiles that are slowly varying, either $\gamma^2$ is small and $\lambda_z$ is large (relative to $r_c$), or $\lambda_z $ is small and $\gamma^2$ is very small, corresponding to extremely shallow ripples on the string. In either case, the ratio $\gamma^2r_c^2/\lambda_z^2$ remains small, so that the rest mass term $\alpha^2$ dominates. We therefore investigate these two limiting regimes separately. Hence,
\begin{eqnarray} \label{dE1}
\Delta E_{|n|} \approx 2\pi\eta^2|n|^2\Delta \times \frac{(2\pi)^2r_{c}^2\gamma^2}{\lambda_z^2} \times \left(\frac{G}{c^4}\right)^{-1}, \ \ \ \alpha^2 \ll \frac{(2\pi)^2r_{c}^2\gamma^2}{\lambda_z^2}.
\end{eqnarray}
\begin{eqnarray} \label{dE2}
\Delta E_{|n|} \approx 2\pi\eta^2|n|^2\Delta \alpha^2 \times \left(\frac{G}{c^4}\right)^{-1}, \ \ \ \ \ \ \ \ \ \ \ \ \ \ \ \  \alpha^2 \gg \frac{(2\pi)^2r_{c}^2\gamma^2}{\lambda_z^2},
\end{eqnarray}
In the first case, we have
\begin{eqnarray} \label{rel}
\Delta E_{|n|} \approx \Delta P_{|n|} c,
\end{eqnarray}
so the emitted energy is mostly in the form of ``radiation", i.e. either gravitational or electromagnetic waves, or massive but highly relativistic particles. At present, it is unclear whether gravitational radiation (see \cite{ViSh:00,HiKi:99} and references therein) or gauge particle emission, which then sources metric perturbations \cite{Vilenkin:1986ku,Bevis:2006mj}, is the primary energy loss mechanism for oscillating Nielsen-Olesen strings. It probable that, whichever mechanism dominates for traveling waves or oscillating loops of cylindrical string, the same mechanism dominates for lumpy strings in this regime. However, the opposite limit, Eq. (\ref{dE2}), we may assume that most of the energy is channeled into the production of (slow moving) massive particles, so that
\begin{eqnarray} \label{nonrel}
\Delta E_{|n|} \approx \Delta M_{|n|} c^2,
\end{eqnarray}
where $\Delta M_{|n|}$ is the total mass of the particles produced, which also depends on the absolute value of the winding number, $|n|.$ \\
\indent
Explicitly writing the correct fundamental constants, the expression for $r_c^2$ in terms of $\eta^2$ and $e^2$ is
\begin{eqnarray} \label{r_c_exp}
r_c^2 \approx \left(\frac{\hbar^2 G}{c^4}\right)(\mu_0 e^2 \eta^2)^{-1}=\frac{1}{4\pi}l_P^2\left(e_r^2\eta ^2\right)^{-1},
\end{eqnarray}
where the Planck length $l_P$ is defined as $l_P^2=\hbar G/c^3$, and $e_r^2=e^2/q_P^2$, where the Planck charge is $q_P^2=4\pi \epsilon _0\hbar c=4\pi \hbar /c\mu _0$. If $e$ denotes the standard charge of the electron, the dimensionless quantitity $e_r$ is equal to the fine structure constant, $\alpha_s \approx 1/137$, though, in principle, we can consider toy models in which the coupling is arbitrary.\\
\indent
We may then estimate the energy emitted by the string section of length $\Delta $, in either scenario above, in terms of $\lambda_z$ and the appropriate numerical constant, $\gamma^2$ or $\alpha^2$, respectively. Using $\Delta =\lambda _z$, this gives
\begin{eqnarray} \label{outputs}
(\Delta E_{|n|})_{rad} \approx 2\pi ^2\frac{\hbar c}{e_r^2}\frac{|n|^2\gamma^2m}{\lambda _z}, \ \ \ (\Delta E_{|n|})_{matter} %\approx \eta^2|n|^2\lambda_z \alpha^2
\approx \frac{1}{2}\frac{\hbar c}{e_r^2}\frac{|n|^2\alpha^2m}{r_c^2}\lambda _z .
\end{eqnarray}
The corresponding power outputs are simply
\begin{eqnarray} \label{power}
\frac{(\Delta E_{|n|})_{rad}}{(\Delta t)} \approx 2\pi ^2\frac{\hbar c^2}{e_r^2}\frac{|n|^2\gamma^2m}{\lambda _z^2}, \ \ \ \frac{(\Delta E_{|n|})_{matter}}{(\Delta t) }%\approx c\eta^2|n|^2 \alpha^2.
\approx \frac{1}{2}\frac{\hbar c^2}{e_r^2}\frac{|n|^2\alpha^2m}{r_c^2}.
\end{eqnarray}
In matter dominated emission, the total energy release is proportional to the lump width whereas, for radiation, it is inversely proportional, as expected intuitively. As a result, the expected power output (per lump) for the former tends to a constant value, whereas, for the latter, it is inversely proportional to the lump width squared. \\
%
%%%%%%%%%%%%%%%%%%%%%%%%%%%%%%%%%%%%%%%%%%%%%%%%%%%%%%%%``
\subsection{Astrophysical applications} \label{Sect.IVB}
Considering a single lump (setting $m=1$) and assuming that $e = 1.602 \times 10^{-9}$ C, the standard charge of the electron, the radiation and matter-dominated power outputs from a string of length $\lambda_z$ can be written in an equivalent form as
\begin{equation} \label{power*}
\frac{(\Delta E_{|n|})_{rad}}{\Delta t}\approx 8.906\times 10^{62}\times |n|^2\gamma ^2\times \left(\frac{\lambda _z}{l_P}\right)^{-2}\;{\rm erg/s}, 
\end{equation}
\begin{equation}  \label{power**}
\frac{(\Delta E_{|n|})_{matter}}{\Delta t } \approx 2.255\times 10^{61}\times |n|^2\gamma ^2\times \left(\frac{r _c}{l_P}\right)^{-2}\;{\rm erg/s}.
\end{equation}
It is interesting to compare the above estimates with the energy emission of superconducting cosmic strings. Superconducting strings, moving through a cosmic plasma in the presence of a magnetic field $\vec{B}$, are sources of synchrotron radiation, which may be observed at centimeter wavelengths. The energy of this radiation can be estimated using the formula \cite{W,VF,AN}
\begin{equation}
\frac{\Delta E_{SS}}{\Delta t}=2\times 10^{32}B_{-6}^{2/3}R_{20}^{5/2}n_p^2\;{\rm erg/s},
\end{equation}
where $R_{20}=R\times 10^{20}\;{\rm cm}$ is the radius of the string loop, $B_{-6}=B/\left(10^{-6}\; {\rm G}\right)$, and $n_p$ is the number density of the cosmic plasma particles. For a typical vacuum situation, $B_{-6} = 10^3$, $R_{20} = 1$ and $n_p = 10^3$ ${\rm cm}^{-3}$, giving $\Delta E_{SS}=2\times 10^{40}$ erg/s. For the case of a very dilute medium, with $R_{20} = 3$, $B_{-6}=1$ and $n_p = 10$ ${\rm cm}^{-3}$, we obtain $\Delta E_{SS}\approx  10^{34}$ erg/s. The radio emission of superconducting strings has a very peculiar structure, in that they should be observed as a closed line.\\
\indent
Thus, for $|n|^2 \sim \gamma^2 \sim \mathcal{O}(1)$, the emitted power of a single-lump string of length $\lambda _z = 10^{44}l_P$, decaying via radiative emission, is greater than or equal to that of a superconducting string of equal length. This implies $\lambda_z \approx 10^{11}$ cm, and the energy is emitted on a timescale $\Delta t \lesssim 1$ millisecond, after the epoch of string formation. However, clearly, our original assumption that radiative emission dominates is invalid for $\lambda_z \gg r_c$, and the latter will be extremely small for any realistic/natural symmetry breaking scale (for example, Planck, GUT or electroweak scales). Nonetheless, as a crude first approximation, we may assume that  the formulae in Eq. (\ref{power*})-(\ref{power**}) hold true, even in the more general case, when the energy is emitted as a mix of massive and relativistic particles.\\
\indent
There are a number of astrophysical observations and phenomena that can be interpreted in the framework of the cosmic string theory. Of these, gamma ray bursts (GRBs) \cite{rev} are one of the most extensively investigated. GRBs are powerful cosmic explosions that can be separated, based on their durations and characteristic wavelengths, in two classes \cite{K}: short duration and hard spectrum, or long duration and soft spectrum, respectively. The possibility that GRBs are related to superconducting string emission via synchrotron radiation was investigated in detail in \cite{Bab}-\cite{Cheng}. By assuming that the emitted energy is focused along a relatively narrow beam, the power released in a GRB event is of the order of $10^{52}$ erg/s \cite{rev}. By assuming a similar beaming mechanism for the radiation emission from lumpy strings, it follows that a standard GRB type explosion can be generated by a single-lump string with $\Delta = \lambda _z\approx 10^{20}l_P\approx 1$ fm. On the other hand smaller cosmic lumpy strings can easily produce outputs of power of the order of $10^{53}-10^{54}$ erg/s, as observed, for example, in the (anomalous) cases of GRB 971214 and GRB 990123 \cite{Kul}. \\
\indent
Lumpy strings may also be a powerful source of matter emission, with the emitted power almost of the same order of magnitude as the radiation power. It has been shown in \cite{Fer}  that a tangle of light, superconducting strings in the Milky Way could be the source of the observed 511 KeV emission from electron-positron annihilation in the Galactic bulge. In principle, powerful positron emission can also take place from lumpy cosmic strings. By assuming that some fraction $\eta _{e^{+}}$ of the total energy is emitted in the form of positrons, we can write the relation $\left(\Delta E_{|n|}\right)_{e^{+}}/\Delta t=\left(\Delta N_{e^{+}}/\Delta t\right)\gamma _Lm_ec^2=\eta _{e^{+}} \left(\Delta E_{|n|}\right)_{matter}$, where $\Delta N_{e^{+}}$ is the number of positrons, $m_e$ is the electron/positron rest mass, and $\gamma _L$ is the Lorentz factor of the positrons. Hence we can estimate the number of the positrons emitted per unit time by the lumpy cosmic strings as
\begin{equation}
\frac{\Delta N_{e^{+}}}{\Delta t}\approx \frac{1}{2\gamma _L}\eta _{e^{+}}\frac{\hbar }{m_ee_r^2}\frac{|n|^2\alpha ^2}{r_c^2},
\end{equation}
or
\begin{equation}
\frac{\Delta N_{e^{+}}}{\Delta t}\approx 2.751\times 10^{58}\times \eta _{e^{+}} \times \left(\frac{\gamma _L}{10^9}\right)^{-1}\left(\frac{r_c}{l_P}\right)^{-2}|n|^2\alpha^2 \  \;{\rm s}^{-1}.%\;{\rm positrons/s}.
\end{equation}
The observed positron production rate in the galactic center is of the order of $\Delta N_{e^{+}}^{(obs)}\approx 1.2\times 10^{43}$ positrons/s \cite{Fer}. By assuming that 1\% of the produced particles are in the form of positrons, $\eta _{e^{+}} =0,01$, the observed positron emission can be explained by the presence of lumpy strings at the galactic center, having $r_c=10^{26}l_P$ (again with $|n|^2 \sim \gamma^2 \sim \mathcal{O}(1)$).  Once produced by the decaying string lumps, the positrons will interact with the interstellar medium. The most common interaction processes are pair annihilation with an ambient electron, and positronium formation \cite{Fer}. As a result of these interactions between positrons and interstellar matter, a narrow 511 keV line results, whose presence has been confirmed by observations \cite{Fer}. 
%
%%%%%%%%%%%%%%%%%%%%%%%%%%%%%%%%%%%%%%%%%%%%%%%%%%%%%%%%
%%%%%%%%%%%%%%%%%%%%%%%%%%%%%%%%%%%%%%%%%%%%%%%%%%%%%%%%``
\subsection{Cosmological applications} \label{Sect.IVC}
However, cosmologically, perhaps the most important implication of the existence of lumpy strings would be their effect on the formation and evolution of string networks. In order to investigate this, detailed models of the field dynamics during the symmetry breaking epoch must be developed in order to determine whether variations in local vortex radii may be correlated over regions $\lambda_z \gg r_c$, as required for macroscopic string sections, filled with propagating lumps, to exist. In principle, this introduces a second ``correlation length" into the network, in addition to the usual distance $d \approx \chi ct$ $(\chi \lesssim 1)$, which determines the average string length \cite{ViSh:00,HiKi:99}. Naively, we may expect this to significantly affect the growth of small scale structure for $\lambda_z \ll d$, smoothing out kinks over a time scale $\Delta t \approx \lambda_z/c$ after the epoch of string formation, or to affect the macroscopic network dynamics for $\lambda_z \gg d$. \\
\indent
In the former case, the initial lumpy strings should evolve into smooth Nielsen-Olesen strings over a time, $\Delta t$, that is insufficient to affect the dynamics of the macroscopic network. In the latter, the effect of lumps propagating on the network is unknown. For small oscillations in the radial profile, lumpy strings may behave just like uncharged current carrying strings, such as chiral strings %\cite{CoHiTu87,Test3,Test4,BlOlVi01} 
\cite{CoHiTu87}-\cite{BlOlVi01} or neutral vortons %\cite{Test1,Test2,Carter(1990),Carter&Martin(1993),Larsen(1993),Martins&Shellard(1998),Carter&Davis(2000),LaWa10,Vortons(2013)} 
\cite{Test1}-\cite{Vortons(2013)} but, for large radial variations, they may resemble necklaces \cite{Berezinsky:1997kd}-\cite{Lake:2009nq} with ``beads" traveling at light speed. \\
\indent
In general, lump propagation should damp traveling wave solutions, such as those presented in \cite{VaVa:90}, when the lump width and amplitude are both smaller than those of the transverse mode, but it would be interesting to investigate the effect of resonance. It may be impossible to excite transverse oscillations, at a point $z$, whose amplitudes are smaller than the local vortex radius, or whose wavelengths are shorter than the lump width. Similarly, lump propagation may help to stabilize loops, of initial length $\rho(t_i) \ll \lambda_z$, over the characteristic time scale for lump decay, but momentum carrying loops of length $\rho(t_i) \lesssim \lambda_z$ could not form, and nor could loops with radii smaller than the \emph{maximum} local string width. Given the possible complexity of all these interactions, it is difficult to estimate what overall effect the presence of lumps would have on the frequency of intersecting strings chopping of from the network, or the ability of the latter to reach a scaling solution \cite{ViSh:00,HiKi:99}. \\
\indent
As an interesting toy model, but perhaps physically pathological case, one could investigate the possibility of local over-densities being sufficiently high to induce gravitational collapse, leading to the spontaneous production of black holes along the string. Accretion rates onto string networks \cite{GonzalezDiaz:2005ex}-\cite{Aguirre:1995ca}, gravitational radiation and lensing properties (see \cite{ViSh:00,HiKi:99} and references therein) and contributions to CMB anisotropies \cite{Miyamoto:2012ck}-\cite{Lazanu:2014xxa} should also be affected by lumps.
%
%For the smallest possible strings containing only one lump, we may set
%\begin{eqnarray} \label{nonrel}
%\Delta = \lambda_z \approx r_c, \ \ \ (m=1),
%\end{eqnarray}
%
%%%%%%%%%%%%%%%%%%%%%%%%%%%%%%%%%%%%%%%%%%%%%%%%%%%%%%%%
%%%%%%%%%%%%%%%%%%%%%%%%%%%%%%%%%%%%%%%%%%%%%%%%%%%%%%%%``
\section{Conclusions}  \label{Sect.V}
We have demonstrated the existence of a new solution of the abelian-Higgs field equations, which describes a generalization of the Nielsen-Olesen string. Physically, the new solution represents a non-cylindrically symmetric field configuration, with variable scalar and vector core radii or, in other words, a lumpy cosmic string. In addition to the standard string parameters, such as the symmetry breaking energy scale $\eta$ and topological winding number $n$, it is characterized by two arbitrary functions, obeying the standard relativistic wave equation, which represent arbitrary left or right movers. For a string with a static central axis, lying along the $z$-direction, these functions determine the local values of the $t$ and $z$-dependent core radii, so that the string carries momentum and the ``lumps" propagate at the speed of light.\\
\indent
Though symmetry breaking from the abelian-Higgs action remains a toy model for string formation, there is every reason to suspect that similar, generalized, solutions may be found for more realistic string species, such as electroweak \cite{Vachaspati:1992uz,James:1992zp,James:1992wb}, or GUT scale strings \cite{Kakushadze:1997mc}. In effect, this introduces an additional phenomenological degree of freedom to the string network, which must be accounted for in any effective action, giving rise to position and time-dependent finite thickness corrections. In principle, there is no maximum value for the local string radius in the classical theory, though further analysis is required to determine whether a limit arises due to quantum effects.\\
\indent
As such, lumpy strings embody several interesting features of existing models simultaneously. If the radial oscillations are highly localized, they resemble necklaces, with highly energetic beads traveling at light speed but, when individual lumps are highly dispersed along the string length, they may be seen as a form of neutral current, propagating in an (approximately) cylindrical, superconducting string. \\
\indent
Since the lumpy string energy density is higher than that of the Nielsen-Olesen string, we expect the field configuration to ``relax" to Nielsen-Olesen form over some characteristic time scale, given approximately by the lump width divided by the propagation speed, i.e. the speed of light. With these simple assumptions, we were able to estimate the power emission from decaying lumps and found that they provide a viable mechanism for explaining anomalous gamma ray bursts \cite{Kul}, in the case of radiation dominated emission, and may contribute to the observed positron excess emanating from the galactic centre  \cite{Fer}, if the decay is dominated by matter emission.
%
%%%%%%%%%%%%%%%%%%%%%%%%%%%%%%%%%%%%%%%%%%%%%%%%%%%%%%%%
%%%%%%%%%%%%%%%%%%%%%%%%%%%%%%%%%%%%%%%%%%%%%%%%%%%%%%%%``
%
\begin{center}
{\bf Acknowledgments}
\end{center}
M.L. wishes to thank Kathy Hung, for correcting his English.
%

%
%\appendix
%
%\section{AppendixA} \label{AppendixA}%``
%
%\textcolor{red}{ADD LATER.}
%
\end{document}